\documentclass[aps, prd, twocolumn, amsmath, amssymb]{revtex4}
\usepackage{graphicx}
\usepackage{color}
\usepackage{natbib}
\begin{document}


\title{Big Bang Nucleosynthesis Constraints on the Self-Gravity of Pressure}
\author{Saul Rappaport}
\affiliation{Department of Physics and Kavli Institute for Astrophysics and 
Space Research, MIT, Cambridge, MA 02139}
\author{Josiah Schwab}
\affiliation{Department of Physics and Kavli Institute for Astrophysics and 
Space Research, MIT, Cambridge, MA 02139}
\author{Scott Burles}
\affiliation{Department of Physics and Kavli Institute for Astrophysics and 
Space Research, MIT, Cambridge, MA 02139}
\author{Gary Steigman}
\affiliation{Departments of Physics and Astronomy and Center for Cosmology 
and Astro-Particle Physics,The Ohio State University, 191 West Woodruff 
Ave., Columbus, OH 43210}


\begin{abstract}
Using big bang nucleosynthesis and present, high-precision measurements 
of light element abundances, we constrain the self-gravity of radiation 
pressure in the early universe. The self-gravity of pressure is strictly 
non-Newtonian, and thus the constraints we set provide a direct test of 
this prediction of general relativity and of the standard,
Robertson-Walker-Friedmann cosmology.
\end{abstract}



\maketitle


\section{Introduction}

Certain aspects of general relativity are well tested.  For example, the 
Schwarzschild metric has been quantitatively verified in the weak-field 
limit on small scales, e.g., the Solar system \citep{Shapiro64, Bertotti03} 
and binary radio pulsars \citetext{e.g., \citealp{Hulse75, Taylor79, Weisberg84}}; 
and on galaxy scales \citep[e.g.,][]{Bolton06}.  In another 
fundamental test of general relativity, the existence of gravity waves has 
been established  \citetext{e.g., \citealp{Taylor82}, \citealp{Weisberg84}}.  
General relativity theory, utilizing the Robertson--Walker metric 
\citetext{\citealp{Friedmann22}, \citealp{Robertson35}, \citealp{Walker36}} 
leads to the Friedmann equations \citetext{\citealp{Friedmann22, Lemaitre27}} 
which govern the expansion behavior of a homogeneous, isotropic Universe. 
However, it is probably fair to say that the Friedmann equations, while 
providing a self-consistent and highly successful framework for cosmology, 
have not been subjected to extensive, independent testing.  In this paper 
we show that one particular aspect of the Friedmann equations, the 
self-gravity of pressure, can be tested quantitatively.

The development of big-bang nucleosynthesis (BBN) codes \citep{Wagoner67, 
Kawano92}, coupled with measurements of the relevant nuclear reaction 
rates \citep{Caughlan88, desc04}, have allowed observations of light element 
abundances to become powerful tools with which to investigate the early 
evolution of the universe. Computational predictions over a wide range 
of parameter space, when compared with primordial abundances inferred 
from observations, have yielded constraints on the current-epoch baryon 
density \citep{Wagoner73, Yang79, KS04, Steigman07}, neutrino physics 
\citep{SSG77, Yang79, KS04, Steigman07}, the fine structure constant 
\citep{Bergstrom99}, the gravitational constant \citep{Steigman76, Yang79, 
Copi04, Steigman07}, primordial magnetic fields \citep{Kernan96}, the 
universal lepton asymmetry \citep{Wagoner67, KS04, Steigman07} and other 
parameters of astrophysical interest. 

Increasingly accurate measurements of element abundances, as well as 
improved understanding of the processes (i.e., stellar and galactic 
nucleosynthesis) which have altered the original abundances, allow 
these restrictions to be continually refined. Deuterium abundances 
\citep{GeissReeves72, Omeara06}, helium abundances \citep{HoyleTayler64, 
IT04, Izotov07}, and lithium abundances \citep{RNB00, Asplund06} have 
all been well measured, although the inferred primordial abundances 
are subject to large and often difficult to quantify systematic 
uncertainties. More recently, observations of the cosmic microwave 
background (CMB) have yielded an independent estimate of $\eta$, the 
baryon to photon ratio at a much later epoch in the evolution of the 
Universe \citep{Spergel07}.

\section{Analysis}
\subsection{Friedmann Equations}

For an isotropically expanding universe in which the matter/energy is 
distributed homogeneously, the expansion of the universe is described 
by a time-dependent scale factor, $a = a(t)$.  In the standard 
Friedmann--Robertson--Walker (FRW) cosmology, the time variation of 
the scale factor is given by the Friedmann equations in terms of the 
average density and pressure.  For example, the ``acceleration'' of 
the scale factor, $\ddot{a}$, 
is given by:
\begin{equation}
\frac{\ddot{a}}{a} = - \frac{4\pi G}{3} \left(\rho + \frac{3P}{c^2}\right) ~,
\end{equation}
where $\rho c^2$ is the energy density and $P$ is the pressure.  This is 
the exact $\mathcal{G}_{rr}$ component of the Einstein field equation for 
a homogenous and isotropic universe.  Note that the $3P$ term, implying 
the self-gravity of pressure, is a purely General Relativistic (GR) effect, 
with no analog in Newtonian gravity. The ``velocity" of the scale factor, 
$\dot{a}$, is given by the Friedmann-Lemaitre equation:
\begin{equation}
\left(\frac{\dot{a}}{a}\right)^2 =  \frac{8\pi G}{3} \rho + \frac{kc^{2}}{a^{2}}~,
\end{equation}
whose origin is the ${\bf \mathcal{G}}_{tt}$ component of the Einstein 
field equations (the second term on the right hand side of the equation 
is due to the curvature; $k$ is the curvature constant which appears in
the Robertson-Walker metric).  For any fluid, given its equation of state, 
i.e., $P = P(\rho)$, eq.~(1) can be integrated to yield eq.~(2) -- but 
only if the $3P$ term is included.  

We would now like to test the Friedmann-Lemaitre equations by placing 
constraints on the existence of the $3P/c^2$ term in eq.~(1), and to 
see what the testable consequences are for eq.~(2).  To do this we will, 
by necessity, no longer be assuming the validity of GR. However, we 
will retain the energy conservation of expanding fluids via the first 
law of thermodynamics (the perfect fluid approximation or entropy 
conservation).  

We start with a ``Newtonian cosmology'' \cite{Milne34, Uzan01} 
which, of course, cannot be completely justified outside the context 
of GR, but which nonetheless provides considerable insight into our testing 
of the $3P/c^2$ term \cite{rindler01}.  For the usual Newtonian gravity, this amounts to 
\begin{equation}
\frac{\ddot a}{a} = - \left(\frac{4\pi}{3} \right) G \rho ~.
\end{equation}
For the special, zero-pressure case where $\rho = \rho_0 a^{-3}$, 
eq.~(3) can be integrated to yield the familiar expression for 
$\dot a$:
\begin{equation}
\left(\frac{\dot{a}}{a}\right)^2 =  \frac{8\pi G}{3} \rho  ~+~ 
\frac{{\rm constant}}{a^2}~.
\end{equation}
For the Newtonian analysis, the constant in eq.~(4) is simply a
constant of integration, in contrast to the curvature term which
appears in the Friedmann equation, eq.~(2).  

This form of eq.~(4) is, however, only valid for the special case of a 
pressureless fluid.  What is missing from eq.~(3) for the general 
case of a fluid with non-zero pressure is a term accounting for the 
self-gravity of pressure (see eq.~1), which has no expression in 
a purely Newtonian formulation.  Suppose we now add such a term 
to eq.~(3), in an {\em ad hoc} fashion, 
with an arbitrary multiplicative constant, $\chi$. 
\begin{equation}
\left(\frac{\ddot{a}}{a}\right) = - \left(\frac{4\pi}{3}\right)  
G  ~\rho \left[1 + \chi(3 P/\rho c^2)\right] ~.
\end{equation}
For $\chi = 1$ we incorporate the full effect of the self-gravity of 
pressure (as it follows from GR; eq.~1), while for $\chi = 0$, this 
non-Newtonian effect is completely neglected.  This is our proposed 
modification of the Friedmann equation, eq.~(1).  Using the first 
law of thermodynamics for an adiabatic expansion expressed as:
\begin{equation}
d\left(\rho c^2 a^3\right) = -P \,d\left(a^3\right)~,
\end{equation}
we can solve eq.~(6) for $\dot a^2$:
\begin{equation}
\dot a^2 = -\frac{8 \pi G}{3} \left\{ (1-3 \chi) \int \rho a~da - 
\chi \int a^2~d\rho \right\} ~.
\end{equation}
Note that only for $\chi = 1$ (i.e., the full implementation of the 
pressure self-gravity term) is the standard form of the $\dot a^2$ 
version of the Friedmann equation recovered, viz,
\begin{equation}
\dot a^2 ~=~\frac{8 \pi G}{3} \int d\left(\rho a^2\right) ~=~  
\frac{8 \pi G}{3} \rho a^2~+~{\rm constant}~.
\end{equation}
For any equation of state of the form $P = w\rho c^2$ where $w$ 
is a constant, the integrals in eq.~(8) yield Friedmann-like 
equations, but with a modified leading coefficient:
\begin{equation}
\left(\frac{\dot{a}}{a}\right)^2 = H^2 =  
\left[\frac{1+3w\chi}{1+3w}\right] \frac{8\pi G}{3} \rho~+\,\frac{{\rm constant}}{a^2}~.
\end{equation}
For a matter-, radiation-, or vacuum energy-dominated universe, 
$3w=3P/\rho c^2 = 0, 1$, and $-3$, respectively.  If $\chi = 1$, the 
proper Friedmann equation is recovered, regardless of the choice of 
equation of state.   If $\chi = 0$, for a pressureless gas (``matter"), 
the correct Friedmann result is recovered; but, as soon as the expanding 
fluid has significant pressure, an incorrect result (with respect to 
GR) is obtained.  Having set aside GR, which would otherwise connect the integration
constant in equations~(8) \& (9) to the geometry, there is no fundamental 
connection to the geometry of the underlying space-time in our proposed 
modification of the Friedmann equations.

For the case where a combination of radiation, matter, and vacuum densities 
are considered together, eq.~(9) can be written in a nearly familiar form:
\begin{widetext}
\begin{equation}
\left(\frac{\dot{a}}{a}\right)^2 =  ~H^2 = ~H_0^2\left[ \frac{\Omega_M}{a^3}  
+\frac{\Omega_R}{a^4} \left(\frac{1+\chi}{2}\right)+\,\Omega_V
\left(\frac{3\chi-1}{2}\right) +\frac{\Omega_k}{a^2} \right]~,
\end{equation}
\end{widetext}
where the $\Omega$'s are defined by $\Omega_j = 8\pi G\rho_{j,0}/3H_0^2$, 
and the $\rho_{j,0}$'s are, in turn, the densities of the respective 
constituents evaluated at the current epoch. Finally, in this expression 
$\Omega_k \equiv 1-\left[\Omega_M + \left({\frac{1+\chi}{2}}\right)\Omega_R 
+ \left({\frac{3\chi-1}{2}}\right)\Omega_V \right]$.  Here we 
reemphasize that $\Omega_k$ can no longer be interpreted in terms of 
the curvature -- it is just a constant of integration.  If $\chi \ne 1$ 
eq.~(10) has consequences for the CMB and current matter/vacuum dominated 
epochs, in addition to the BBN epoch, some of which could independently 
constrain $\chi$, but these are beyond the scope of the present work.

We now focus on the epoch when the energy density contributed by relativistic
particles (including relativistic neutrinos), ``radiation", dominated 
the energy density of the Universe and radiation pressure is important.  
Our goal is to constrain $\chi$ by comparing BBN predictions with 
observations of light element abundances. The expansion rate of the 
Universe during this epoch is described by
\begin{equation}
\left(\frac{\dot{a}}{a}\right)^2 =   \left(\frac{1+\chi}{2}\right)~\frac{8\pi G}{3}~\rho_R~,
\end{equation}
where the $\Omega_k/a^2$ term which appears in eq.~(10) has been dropped 
because it is negligible compared to the $\Omega_R/a^4$ term during the 
radiation-dominated, BBN epoch.  As revealed by eq.~(11), the effect of a
value of $\chi$ which differs from unity is to change the early-universe
expansion rate (Hubble parameter) from its standard value.  In this sense,
$\chi \neq 1$ is equivalent to an early-universe value of the gravitational
constant \citep{Steigman76} which differs from its present value or, a total 
relativistic energy density which differs from its standard-model value 
(as often parameterized by the effective number of neutrinos: 
$\rho_{R}'/\rho_{R} \equiv 1 + 7\Delta$N$_{\nu}/43$)~\citep{SSG77}.

\subsection{Nucleosynthesis Calculations}

Nucleosynthesis calculations were performed with a BBN code which has been 
updated with the latest reaction rates and whose output has been compared
to that of other, published codes.  Since the parameter we seek to constrain, 
i.e., $(1+\chi$)/2, is multiplicative with $G$, we have simply varied $G$ as 
a surrogate for $\chi$.  Thus, in our case the BBN-predicted abundances are
functions of the baryon density parameter $\eta_{10} = 10^{10}\eta \equiv 
10^{10}(n_{\rm B}/n_{\gamma})$ and $G$. In this work $\sim$200000 BBN 
calculations were performed, varying $\chi$ from 0 to 2 in steps of 0.002 
(or equivalently $G/G_0$ from 1/2 to 3/2 in linear steps of 0.001) and 
varying $\log \eta$ from ${-10}$ to ${-9}$ in steps of 1/200 dex. The 
results of these calculations are the isoabundance contours for deuterium, 
helium-4, and lithium-7 shown in the $\{\eta$, $\chi\}$ plane in Figure\.~1.
As estimated in \cite{Burles01b}, for fixed values of $\eta_{10}$ and 
$G$, the uncertainties in the nuclear reaction rates contribute a $\sim 
3$\% uncertainty ($\sim 1\sigma$) to the BBN-predicted abundance of 
deuterium, and as estimated in \cite{Burles01a} $\gtrsim $~0.2\% ($\sim 
1\sigma$) for the $^{4}$He mass fraction. 

\begin{figure*}[t]
\centering
\includegraphics[width = 6.5in]{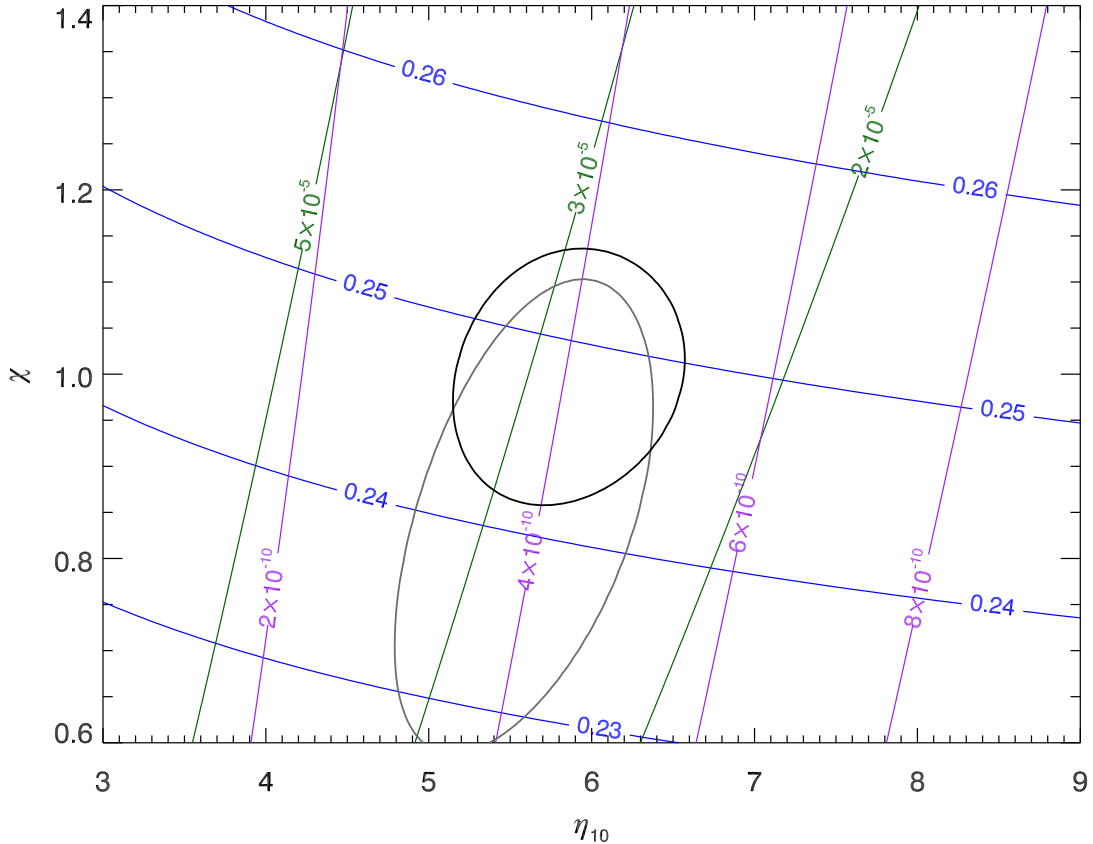}
\caption{Isoabundance contours for the predicted primordial abundances of 
deuterium and helium-4 in the $\chi - \eta_{10}$ plane, along with 90\% 
contours for a choice of the primordial D abundance and two choices for 
the primordial $^4$He abundance (see the text for details). Green contours 
are for deuterium abundances (D/H)$_P$. Blue contours are for the $^4$He 
mass fraction Y$_P$.  The thick, black ellipse is the 90\% confidence 
contour using the \citet{PLP07} helium-4 abundance, and the thin, grey 
error ellipse is the 90\% confidence contour corresponding to the more 
conservative Steigman \cite{Steigman07} helium-4 abundance.  The purple 
contours are for the (Li/H)$_P$ abundance; however, these were not used 
to constrain either $\chi$ or $\eta$.}
\end{figure*}

\subsection{Light Element Observations}

Deuterium provides an excellent constraint on the baryon density because 
its post-BBN evolution is simple (D is only destroyed when gas is cycled 
through stars) and the observed amounts require that it must have formed 
in the big bang rather than in stellar or galactic processes \citep{Reeves73, 
els76}.  Also, its BBN-predicted abundance is extremely sensitive to the 
baryon to photon ratio, $D/H \propto \eta^{-1.6}$ \citep{Yang79, Burles01a}. 
Deuterium measurements along the lines of sight to high redshift quasars 
have led to the current determination of $\log (D/H)_P = -4.55 \pm 0.04$ 
\citetext{$1\,\sigma$ confidence; \citealp{Omeara06}}.  Contours of constant 
deuterium abundance (by number) are shown as green curves in Fig.\,1.

Although BBN production of $^4$He is relatively insensitive to $\eta$, 
its abundance provides an extremely useful constraint on the early-universe
expansion rate (the Hubble parameter) and, therefore, on $\chi$. As stars 
and galaxies evolve, stellar nucleosynthesis results in some post-BBN 
production of $^4$He.  As a consequence, the primordial abundance (mass
fraction) of $^4$He, Y$_P$, is best determined from present-day observations 
of low-metallicity, extragalactic HII regions which are less contaminated 
by post-BBN produced $^4$He. Since the total number of such HII regions 
exceeds 80 \citep{IT04}, it is not surprising that the formal, statistical 
uncertainty in Y$_P$ is small.  However, it has been well known for 
decades \citep{DK85} that systematic corrections, such as underlying 
stellar absorption, ionization corrections, collisional excitations, 
etc., have the potential to change the central value of Y$_P$ as well 
as to increase significantly the error budget.  The largest data set 
of consistently observed and analyzed HII regions is from \citet{IT04} 
who find Y$_P = 0.243 \pm 0.001$.  Izotov et al. have recently revised
this to $0.247 \pm 0.001$ \citetext{$1\,\sigma$ confidence; \citealp{Izotov07}}.
These analyses largely ignore most sources of systematic uncertainty, 
resulting in an error, largely statistical, which is too small to 
reflect the true uncertainty in Y$_P$.  Accounting for some, but not 
all sources of systematics, and employing a model-dependent linear 
extrapolation of Y to zero oxygen abundance, Y$_P$ has very recently 
been inferred by \citet{PLP07} to be $0.248 \pm 0.003$.  Contours of 
constant helium-4 mass fraction are shown as blue curves in Fig.\,1.

It is interesting to note that for SBBN ($\chi = 1$), as well as for 
BBN with $\chi$ allowed to be free, the predicted primordial abundances 
of deuterium and lithium are strongly coupled~\citep{KS04}; see, Fig.~1.  
For our choice of the primordial D abundance, and for either choice of 
the primordial $^4$He abundance, the predicted primordial lithium 
abundance lies in the range 12+log(Li/H) = 2.6 -- 2.7.  This is in 
contrast to the best determinations of the lithium abundance in the 
oldest, most metal-poor stars in the halo of the Galaxy, where 
12+log(Li/H) $\simeq 2.1$~\citetext{\citealp{RNB00, Asplund06}}.  The 
generally accepted explanation of this factor of 3 -- 4 discrepancy 
is that the lithium observed at present in these oldest stars in the 
Galaxy has been diluted/depleted from the initial lithium abundance 
in the gas out of which these, nearly primordial, stars formed 
\citetext{\citealp{Pinsonneault02, Korn06}} but, for a contrary point 
of view, see \citet{bonifacio07}.

\subsection{BBN Constraints on $\chi$}

Figure 1 displays the results of our analysis. A section of the $\{\eta$, 
$\chi\}$ parameter space is shown, with {\em number} density (relative 
to hydrogen) contours for deuterium shown in green, for lithium shown 
in purple, and for the {\em mass} fraction of He-4 shown in blue.  Notice 
that the pairs of $\{$D/H, Y$_P\}$ or of $\{$Li/H, Y$_P\}$ abundances form 
nearly orthogonal grids in the $\{\eta$, $\chi\}$ plane, so that the 
primordial abundances of either pair of these nuclides are sufficient 
to bound the cosmologically interesting parameters $\chi$ and the baryon 
density parameter $\eta_{10}$.  Given the uncertainty in inferring the 
primordial lithium abundances from the observational data, only the 
deuterium and helium-4 pair is used in our analysis.  

Assuming statistically-independent Gaussian errors (almost certainly, 
neither the errors in D nor those in $^4$He are truly Gaussian), one 
can calculate the probability, via a maximum likelihood analysis, that 
the abundance determinations agree with the corresponding results of 
the BBN calculations at a given point in the $\{\eta$, $\chi\}$ parameter 
space.  The thick black contour is for the 90\% range in $\eta_{10}$ 
and $\chi$ corresponding to the \citet{Omeara06} deuterium abundance 
and the \citet{PLP07} helium abundance.  The narrowness of this contour 
in the vertical ($\chi$) direction is a direct consequence of the size 
of the \citet{PLP07} estimate of the error in Y$_P$.  Given the sensitivity 
of $\chi$ to Y$_P$, it is interesting to explore the consequence of 
adopting a different central value and uncertainty in Y$_P$, while 
keeping the same primordial D abundance.  To this end, we choose 
Y$_P = 0.240 \pm 0.006$ from \cite{Steigman07}. The thin, gray contour 
in Fig.\,1 corresponds to the 90\% range for this alternate choice of 
Y$_P$.  Both choices are consistent with the standard, Friedman-Lemaitre
result $\chi = 1$.

\section{Conclusions}

As illustrated in Figure 1, the combined constraints are, within the 
uncertainties, consistent with the general relativity prediction of 
$\chi = 1$ and the independent (of BBN) WMAP constraint on $\eta_{10}$ 
of $6.1 \pm 0.2$ \citep{Spergel07} which corresponds to $\chi = 1$.  For 
the \citet{PLP07} choice of Y$_{P}$, $\chi = 1.00\pm 0.14$, while for the 
Steigman \cite{Steigman07} helium abundance, $\chi = 0.84\pm 0.25$.  Note
that the data strongly exclude $\chi = 0$.  The current light element 
observations and BBN computations have provided a test of the general 
relativistic self-gravity of pressure.  Since the modification of GR 
we are testing corresponds, for the radiation-dominated evolution 
appropriate for BBN, to an overall multiplicative factor of the 
product of Newton's gravitational constant and the radiation density, 
$G\rho \rightarrow G\rho \left({\frac{1 + \chi}{ 2}}\right)$, our result 
is equivalent to the BBN constraint on the variation of Newton's constant 
or, alternatively, to a modification of the radiation energy density as 
parameterized by the effective number of neutrinos (see \S1 for references).
\begin{equation}
\frac{1 + \chi}{2} = 1 + \frac{\Delta G}{G} = 1 + \frac{7\Delta N_{\nu}}{ 43}
\end{equation}
Assuming that these other parameters take on their standard-model 
values ($\Delta G = \Delta N_{\nu} = 0$), the self-gravity of the 
radiation (photons and neutrinos) pressure during the BBN epoch has 
been constrained quantitatively.

\begin{acknowledgments}
JS acknowledges support from the Paul E. Gray (1954) Endowed Fund 
for UROP.  The research of GS is supported by a grant from the DOE.  
We thank Ed Bertschinger, James Felten, Alan Guth, Scott Hughes, 
and Alan Levine for extremely helpful discussions.
\end{acknowledgments}





\end{document}